# THE DESIGN OF A SPACE-BASED OBSERVATION AND TRACKING SYSTEM FOR INTERSTELLAR OBJECTS

Ravi teja Nallapu,[*] Yinan Xu,[†] Abraham Marquez,[‡] Tristan Schuler,[§] and Jekanthan Thangavelautham[**]

The recent observation of interstellar objects, 1I/ 'Oumuamua and 2I/ Borisov cross the solar system opened new opportunities for planetary science and planetary defense. As the first confirmed objects originating outside of the solar system, there are myriads of origin questions to explore and discuss, including where they came from, how did they get here and what are they composed of. Besides, there is a need to be cognizant especially if such interstellar objects pass by the Earth of potential dangers of impact. Specifically, in the case of 'Oumuamua, which was detected after its perihelion, passed by the Earth at around 0.2 AU, with an estimated excess speed of 60 km/s relative to the Earth. Without enough forewarning time, a collision with such high-speed objects can pose a catastrophic danger to all life Earth. Such challenges underscore the importance of detection and exploration systems to study these interstellar visitors. The detection system can include a spacecraft constellation with zenith-pointing telescope spacecraft. After an event is detected, a spacecraft swarm can be deployed from Earth to flyby past the visitor. The flyby can then be designed to perform a proximity operation of interest. This work aims to develop algorithms to design these swarm missions through the IDEAS (Integrated Design Engineering & Automation of Swarms) architecture. Specifically, we develop automated algorithms to design an Earth-based detection constellation and a spacecraft swarm that generated detailed surface maps of the visitor during the rendezvous, along with their heliocentric cruise trajectories. The constellation is designed as an optimal zenith-pointing Walker-Delta constellation that meets a specified detection success rate, despite being subjected to pointing constraints and random spacecraft outages. The heliocentric trajectories of the spacecraft swarm are then designed as optimal Lambert arcs that meet launch and arrival requirements. Finally, the operations of swarm around the visitor are optimized to meet a coverage requirement specified by the mission designer. A crucial challenge faced while studying the spacecraft coverage arises from the tumbling dynamics of the visitor. Additionally, the uncertainty in the spin axis of these objects, and their non-spherical shapes prohibit the use of deterministic coverage modeling algorithms. To address these challenges, we develop a new method to study the spacecraft coverage, called the dual-sphere method, where the irregular body is decomposed into two spheres to compensate for its range and field of view. We then optimize the swarm trajectories that statistically meet the coverage requirement using a Monte-Carlo simulation over the uncertainties. Finally, the algorithms described are demonstrated by designing a notional mission to detect and map 1I/ 'Oumuamua, assuming there was enough warning time, using the IDEAS architecture.

## INTRODUCTION

The transit of the interstellar visitor 1I/ 'Oumuamua[1] (shown in Figure 1) through the solar system in October 2017, market the first confirmed observation of an interstellar object originating outside our solar system. Its observation sparked interest in both the planetary science and planetary defense communities. Information on parameters such as composition, origin, size, and shape is of great value to the scientific community. As such the time-limited observations of the visitor still had several uncertainties regarding these parameters[2]. Additionally, the perigee of 'Oumuamua is estimated to be around **0. 2 AU** which occurred in October 2017[3]. During its perigee, the speed of the visitor is estimated to be about **60 km/s** with respect to the Earth. A collision at these high speeds can be catastrophic to life on Earth. While the uniqueness of these visitors crossing the solar system was still being debated, the detection of the second interstellar comet **2I/ Borisov**[4] (shown in Figure 1) conclusively suggested that given time more transits from such visitors can be expected. Specifically, in the case of 'Oumuamua, the first detection was nearly **40** days past its perihelion[5,6], and some of the first reported ephemeris of it was generated about a month past its detection.

---

[*] Ph.D. Candidate, Department of Aerospace and Mechanical Engineering, University of Arizona.
[†] Undergraduate Researcher, Department of Aerospace and Mechanical Engineering, University of Arizona.
[‡] Undergraduate Researcher, Department of Aerospace and Mechanical Engineering, University of Arizona.
[§] Graduate Researcher, Department of Aerospace and Mechanical Engineering, University of Arizona.
[****] Assistant Professor, Department of Aerospace and Mechanical Engineering, University of Arizona.

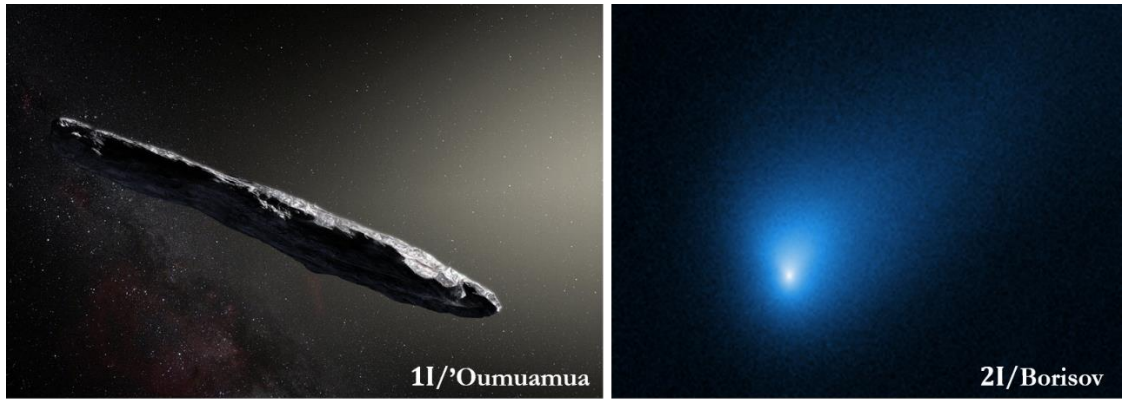

**Figure 1. Confirmed interstellar visitors that have transited the solar system: artistic rendering of 1I/'Oumuamua (left) and 2I/ Borisov (right)   [image source: nasa.gov].**

These limitations underscore the importance of realizing a detection system which provides us enough fore-warning time to design rendezvous missions to study these objects in close detail. Clearly such missions: detection, and rendezvous, are quite complex for a single spacecraft mission. This work focuses on the development of algorithms to design such missions to detect and explore these visitors, using a swarm of spacecraft. The rendezvous missions considered in this work will be visual mapping flyby missions, that generate detailed surface maps of the target body upon close arrival. The important challenges faced by such designs can be described as follows:

In the case of designing a detection swarm, the randomness in the arrival times of these visitors needs to be considered. In such cases, the swarm should be designed to meet a detection probability. Additionally, realistic constraints such as spacecraft outages, and Sun constraints should be factored into the design. Upon detection, the Earth to visitor trajectories of the swarm must be designed to meet practical constraints such as the launch energy, and encounter velocities. Finally, the proximity operations of the swarm upon rendezvous must be optimized to provide optimal visual coverage with a minimal number of spacecraft in the swarm. However, the proximity operation design should also be handled in a stochastic manner due to three important reasons: First, the design must accommodate the fact that the visitor can be in a tumbling state. Second, the design should accommodate the fact that visitors have irregular shapes. Finally, the imaging operation of the swarm can only be performed on the Sun-illuminated side of the visitor. Failure to address these constraints can lead to designs that can guarantee mission success for some attitude configurations (attitude, and angular velocity) of the visitor, while failure in others. Clearly, such designs are complex multi-disciplinary problems to be handled by traditional mission design approaches.

In this work, we develop swarm design algorithms that address the above-mentioned challenges using the IDEAS (Integrated Design Engineering and Automation of Swarms) software, which was developed to design simulated spacecraft swarm mission design to explore solar system small bodies.

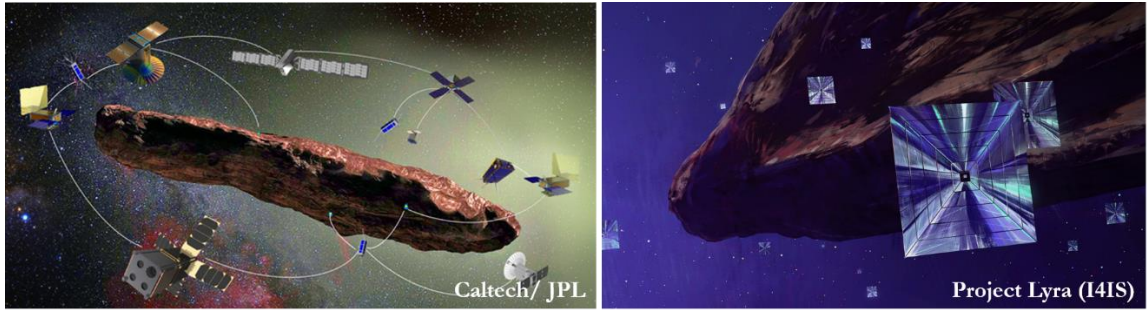

**Figure 2. Conceptual swarm designs to explore interstellar visitors: A swarm with distributed sensing architecture (left) and a rendezvous swarm with laser sails (right) [image sources listed in the picture].**

The detection swarm is designed as a Walker-Delta constellation which is optimized to meet the specified detection probability despite limitations such as random spacecraft outages, and solar pointing constraints. Next, the interplanetary trajectories of the swarm are described as an optimal Lambert arc search problem, where heliocentric trajectories that meet the specified launch energy, and arrival velocity constraints are satisfied. Following this, the visual mapping proximity operation is modeled using a set of Monte-Carlo simulations over multiple designs. Here we present a new method called the dual sphere method that eliminates the sensitivities of the designs to the initial attitude of the visitor. Essentially, the method studies the coverage of the designed swarm over two spheres generated from the maximum and minimum radius of the visitor's shape, allowing us to study the statistical coverage of each swarm design. Finally, the algorithms described in this work are demonstrated through numerical simulations of designing a detection swarm and rendezvous swarm missions to 1I/ 'Oumuamua, assuming enough forewarning time.

The organization of the current work is as follows. The next section describes the related work regarding interstellar visitors, and swarm mission designs. Following this, we present the modeling algorithms used in the current study. Here we present the constellation design, and coverage modeling algorithms, along with the optimization problems that will be solved using the IDEAS architecture. We then proceed to a demonstration of these algorithms using numerical simulations. Here we design the optimal detection constellation to detect these visitors in advance, and then move onto designing the optimal rendezvous swarm. Finally, we conclude the work by summarizing the current work and identifying future work in advancing spacecraft swarm technologies through the IDEAS architecture.

**RELATED WORK**

The detection of 1I/ 'Oumuamua (previously designated U1 'Oumuamua) in October 2017 has opened a new class of small bodies to be explored in the solar system: Interstellar visitors[7]. However, due to the limited resources available, the information available on these objects is highly limited[2]. The object is classified as an asteroid[8], while some cometary behavior was hypothesized[9] during its perihelion. The visitor 2I/ Borisov, on the other hand, exhibited confirmed cometary behavior such as the cometary dust[10]. Of special interest in designing missions to such bodies are its shape and dynamics. Currently, 'Oumuamua is estimated to roughly be a triaxial ellipsoid[11] with its semi-axes lengths as $230 \times 35 \times 35$ m. The object is assumed to be tumbling with a period of about $7.5 - 8.25$ hrs, while the spin axis is not precisely known[5,12,13]. However, studies seem to suggest that the object is in the long axis mode (LAM) of rotation[14] and is expected to return to a uniform major axis spin in about $10^9$ yrs[15]. Due to the unique nature of these objects, they have been the topic of active research in the field of spacecraft engineering. Research has spanned from workshops that focused on technology development of swarms to explore such objects[16], to specific mission designs to these objects[17]. The institute for interstellar studies (I4IS) has studied the feasibility of rendezvous of spacecraft swarms with 'Oumuamua[18] using gravity assists, and laser sailing spacecraft[19]. Additionally, the feasibility of a flyby mission to the comet Borisov has also been studied[20].

Our work has focused on developing tools and algorithms to design such swarm flyby missions to small bodies. We have developed the Integrated Design Engineering and Automation of Swarms (IDEAS) software to provide a unified platform to design multidisciplinary optimal swam missions. In our previous work, we presented the IDEAS architecture, where a swarm mission design problem is decoupled into three sub-design problems (trajectory design, swarm design, and spacecraft design), here we also presented a new classification of spacecraft swarms ranging from constellations to formation flying missions. The visual coverage algorithms for these swarms were presented using the linear camera transformation at the instantaneous location of the spacecraft. The algorithms were demonstrated using visual mapping tasks on uniformly rotating asteroids [21, 22]. Following this, we advanced the capability of IDEAS to design flyby missions to planetary moons using co-orbits around their central planet[23], and direct flybys[24]. Additionally, the principles of design employed in the IDEAS framework were used to design swarms such as a Walker-Delta constellation[25, 26] for meteor detection in low Earth orbits[27, 28] (LEO), and a relay constellation in the Earth-Moon system[29] using Halo orbits around their colinear Lagrange points[30]. The current work focuses on developing statistical tools to design interstellar object detection constellations and visual mapping missions to tumbling objects in the presence of parametric uncertainties.

## METHODOLOGY

This section presents the design methodology used in the current work. We begin by describing the detection swarm design, and its requirements. Next, we proceed to design the heliocentric cruise for a directly launched swarm and then proceed to discuss the visual mapping operation around the target visitor.

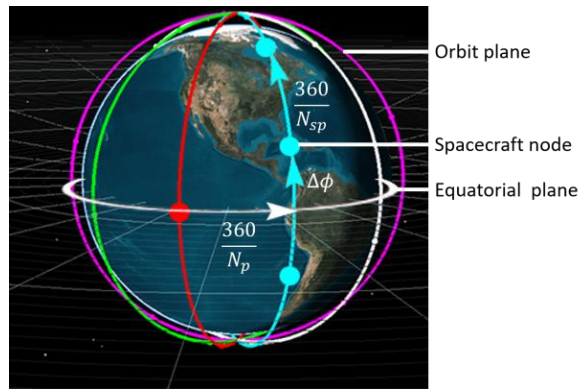

**Figure 3. The geometry of a Walker-Delta constellation of the pattern 90**: $N_C/N_p/F$ **showing the significance of different design parameters.**

### Detection Swarm Design

As mentioned above, we will consider a space-based detection system comprising of a swarm of telescope spacecraft in low Earth orbits (LEO). The swarm will be arranged in a Walker-Delta constellation. The constellation will be generated from a seed spacecraft in a circular orbit with an inclination $in_{sc}$. The orbit of the seed spacecraft is then uniformly divided into $N_{sp}$ segments, which identify the true anomalies of the $N_{sp}$ in-plane spacecraft in the constellation. The orbital plane is then used to define $N_p$ identical orbital planes which have a uniformly distributed right ascension of the ascending node (RAAN). The total number of spacecraft in the constellation is therefore given by

$$N_C = N_p N_{sp} \qquad (1)$$

However, in order to avoid collision between spacecraft in the neighboring plane, a phasing walker parameter $F$ is specified, which specifies the phasing angle $\Delta\emptyset$ used to rotate each neighboring orbital plane as

$$\Delta\emptyset = F \frac{2\pi}{N_C} \qquad (2)$$

The design of a Walker constellation is specified by its design parameters as $in: N_C/N_p/F$. The geometrical parameters of the Walker-Delta constellation are presented in Figure 3. In this work, the orbit of seed spacecraft is selected from repeat ground-track (RGT) orbits, which can ensure that the orbital ground tracks repeat within $N_{orb}$ spacecraft orbits or $N_{rot}$ rotations of Earth[25, 30].

*Attitude Behaviors and Constraints.* The spacecraft in the constellation are assumed to be consisting of zenith-pointing spacecraft, i.e., the spacecraft are assumed to point radially outwards out of their geocentric orbit. However, in order to prevent detector saturation from the Sun, only events detected outside 4 half-solar angles. with respect to the spacecraft are considered as successful detection as shown in Figure 4. In order to realize this, we define a geocentric coordinate system, with $x$-axis along the Earth-Sun line and $z$-axis along the plane normal to the ecliptic.

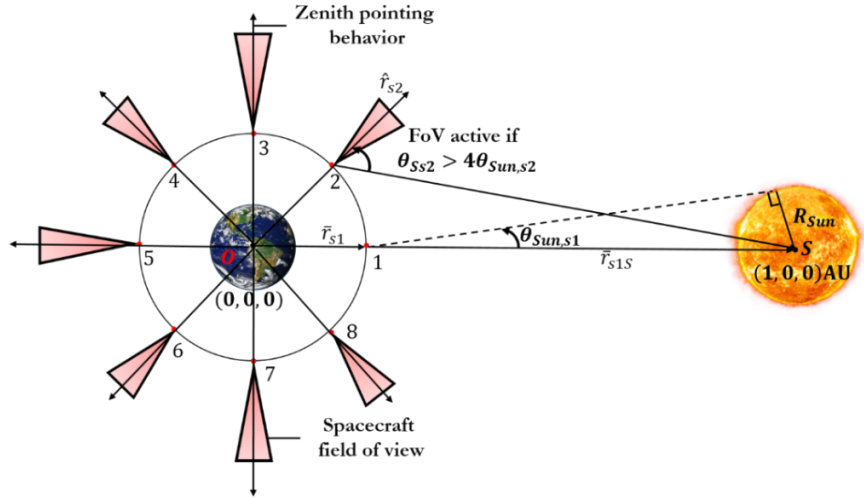

**Figure 4. Attitude behaviors and constraints of the zentith pointing detection constellation.**

The Sun can be located at $\bar{r}_S = [1 \quad 0 \quad 0]^T$ AU. Upon generating a constellation, the geocentric positions of spacecraft $i$ in the swarm $\bar{r}_{si}$ is obtained from its orbital elements[30]. The relative line of sight (LoS) between the spacecraft and the Sun $\bar{r}_{siS}$ can be obtained by the difference between these position vectors. Thus, a pointing constraint can be placed such that the field of view (FoV) of a spacecraft is active, only if the angle between zenith direction $\hat{r}_{si}$, and the spacecraft-Sun LoS is at least 4 times greater than the half solar eclipse angle, i.e.

$$\cos^{-1}\left(\frac{\hat{r}_{si}^T \bar{r}_{siS}}{|\bar{r}_{siS}|}\right) > 4\sin^{-1}\left(\frac{R_{Sun}}{|\bar{r}_{siS}|}\right) \qquad (3)$$

Where $R_{Sun}$ is the radius of the Sun. The constraint in Equation 3 ensures that the spacecraft can only observe a region of the space that is at least two FoV away from the Sun. Additionally, similar constraints on the Earth is placed in order to avoid eclipsing from the Earth.

*Constellation Outages.* When active over long periods of time, it is a reasonable assumption to expect that certain spacecraft in the constellation will be defunct due to random outages. We factor this into the design using a parameter $P_{Op}$ which indicates the percentage of operational spacecraft in each design. Since we employ a Monte-Carlo simulation to study the probability of detection, each constellation is modeled with $N_{inop}$ randomly selected spacecraft which are inoperative. The number of inoperative spacecraft is given by

$$N_{inop} = \text{ceil}\left(\left(\frac{100 - P_{Op}}{100}\right) N_C\right) \quad (4)$$

*Detection Criterion.* Each constellation design is subjected to $N_{mon1}$ Monte-Carlo simulations where an interstellar visitor is generated at random. The event is generated using its spherical coordinates: geocentric radius $r_V$, right ascension $RA_v$, and declination angle $De_v$. The radius $r_V$ is uniformly generated within $[0.7, 1.5]$ AU, the right ascension angles are uniformly generated within $[0, 2\pi]$ rad and the declination angles are generated uniformly in $\left[-\frac{\pi}{2}, \frac{\pi}{2}\right]$ rad. The bounds on the radius are selected such that we can detect events within $0.7 - 1.5$ AU, which can plausibly present enough forewarning window to design a rendezvous mission. In order to receive confirmation, we place a requirement that the event must be detected by at least two spacecraft. The FoV of spacecraft is simulated with a half-FoV of about 30 deg. The generated asteroids are then subjected to the linear camera transformation, which transforms a point on the physical space to the image space of the camera. The event is inside the FoV of a spacecraft if the image transformed event lies within a unit cube[31]. Finally, for a simulated visitor event, the design trial is said to be successful if the visitor falls inside the active FoV of at least two functional spacecraft. The detection efficiency of the constellation with an operational percentage of $P_{Op}$ can now be defined as

$$P_{suc}|_{P_{Op}} = \left(\frac{N_D}{N_{mon1}}\right) \times 100\ \% \quad (5)$$

Where $N_D$ is the number of simulations where the constellation was able to successfully detect the simulated visitor event.

*Constellation Design.* We can now pose the optimization problem solved by the Swarm Designer module of IDEAS to obtain the optimal constellation as

$$\min J_C = N_C = N_p N_{sp}$$

such that

$$P_{suc}|_{P_{Op}} \geq P_r \quad (6)$$

Where $P_r$ is the required detection efficiency specified by the designer. The gene map of the constellation design problem indicating the corresponding design parameters are presented in Figure 5.

| Parameter | # spacecraft orbits | # Earth days | Inclination | # planes | # spacecraft per plane | Spacing parameter |
|---|---|---|---|---|---|---|
| | Seed spacecraft | | | Walker constellation | | |
| Variable | $N_{orb}$ | $N_{rot}$ | $in_{sc}$ | $N_p$ | $N_{sp}$ | $F$ |
| Range | Integer $[1, N_{1,max}]$ | Integer $[1, N_{2,max}]$ | Real $[0, \frac{\pi}{2}]$ | Integer $[1, N_{3,max}]$ | Integer $[1, N_{4,max}]$ | Integer $[1, N_{5,max}]$ |

(The first two columns under Seed spacecraft form the RGT orbit.)

**Figure 5. Gene map listing the design parameters of the constellation design problem.**

**Trajectory Design**

Upon detection with enough forewarning time, a swarm can be launched from Earth to rendezvous with the visitor. The trajectory design problem can be posed as the standard Lambert's problem where the starting location, destination, and time of flight are specified, and the terminal velocities at the start and end of the trajectory are solved. If the force acting on the spacecraft is assumed to be arising primarily due to a single spherical central body, then a Lambert's solver can be used to obtain these terminal velocities[30]. If additional perturbations need to be factored in, a single shooting differential correction scheme can be used to solve the same problem. For designing interplanetary trajectories Lambert's solvers can be used to provide the initial trajectory as the spacecraft will largely be under the gravitational influence of the Sun for most of the trip. The spacecraft in the rendezvous swarm are assumed to be originating at Earth, whose Ephemeris is assumed to follow a Julian date-based regression model[30]. The destination location is assumed to be the target visitor. The ephemeris of the visitor is noted from an epoch and is used to compute its location on any other date[30] assuming the spherical-Sun gravitational model. Thus, under the two-body gravitational dynamics due to the Sun, the trajectory design problem can be reduced to finding a launch date $D_L$ from the Earth, and the required time of flight $ToF$ to arrive at the visitor as shown in Figure 6. The trajectory solvers can then compute the hyperbolic excess velocities during Earth departure $\bar{V}_{\infty,E}^+$ and arrival at the target $\bar{V}_{\infty,T}^-$. The excess velocities identify the critical mission design parameters such as the launch energy

$$C_{3,E} = \left|\bar{V}_{\infty,E}^+\right|^2 = v_{\infty,E}^2 \tag{7}$$

and magnitude of excess velocity at arrival $v_{\infty,T} = \left|\bar{V}_{\infty,T}^-\right|$. Since the trajectories will lead to encounters with a hyperbolic target, we can place a requirement that both the launch energy $C_{3,E}$ and the magnitude of excess velocity on arrival $v_{\infty,T}$ are bounded by $C_{3,max}$ and $v_{\infty,max}$ respectively, which indicate practical limitations by the launch provider, and spacecraft designer.

*Trajectory Design.* We can now pose the optimization problem solved by the Trajectory Designer module of IDEAS to obtain the optimal trajectory as

$$\min J_{Tr} = \left(\frac{C_{3,E}}{C_{3,max}}\right) + \left(\frac{v_{\infty,T}}{v_{\infty,max}}\right)^2$$

such that

$$\begin{aligned} C_{3,E} &\leq C_{3,max} \\ v_{\infty,T} &\leq v_{\infty,max} \end{aligned} \tag{8}$$

| Parameter | Launch date | Time of Flight |
|---|---|---|
| Variable | $D_L$ | $ToF$ |
| Range | Real $[D_{L,min}, D_{L,max}]$ | Real $[ToF_{min}, ToF_{max}]$ |

**Figure 6. Gene map listing the design parameters of the trajectory design problem.**

**Mapping Swarm Design**

The mapping operation describes the proximity operation of the swarm upon their close encounter with the visitor. The goal of the swarm is to generate a detailed surface map of $P_{Cov}$ percentage of the visitor's surface at a required ground resolution $x_{Des}$, and an observation elevation angle of $\varepsilon_{Des}$ using a minimum number of spacecraft. The required flyby altitude for such a flyby can be determined as[26]

$$h_{max} = \left(\frac{x_{Des}}{\lambda}\right) D_{sc} \tag{9}$$

Where, $D_{sc}$ is the aperture diameter of the spacecraft camera, and $\lambda$ is the wavelength of the imaging spectrum. The red spectrum can be used to estimate the upper bound on the imaging altitude. Using a tolerance parameter $\Delta h$, the flyby can be designed to occur at an altitude of $h_F$ given by

$$h_F = h_{max} - \Delta h \tag{10}$$

The half field of view required by the spacecraft $\eta_{sc}$ at this altitude to satisfy the elevation angle constraint is given by[25]

$$\sin \eta_{sc} = \cos \varepsilon_{Des} \sin \theta_H \qquad (11)$$

Where $\theta_H$ is the half-horizon angle of the target body with respect to the spacecraft location.

*Swarm Design Space.* The swarm is assumed to visit the target visitor in $N_v$ visits. Each visit is specified by time past the designed arrival $T_j$. As such, all visits can be designed to occur within $N_{RT}$ rotation periods of the visitor $T_P$. During the $j^{th}$ visit, the visitor is assumed to be visited by $N_{v,j}$ spacecraft. Therefore, the number of spacecraft in the swarm is given by

$$N_{SW} = \sum_{j=1}^{N_v} N_{v,j} \qquad (12)$$

The rendezvous location of the spacecraft $i$ in the swarm, during its closest approach, is specified by an inertial spherical coordinates $r_{map}$, $\theta_{x,i}$, and $\theta_{y,i}$ with respect to the visitor body. The radius $r_{map}$ can be designed based on the specifications of the spacecraft camera. The right ascension angle $\theta_{x,i}$, and $\theta_{y,i}$ as a result of the swarm optimization. The design space of the mapping swarm design problem of an example swarm is presented in Figure 7 and the gene map of the design variables is presented in Figure 8.

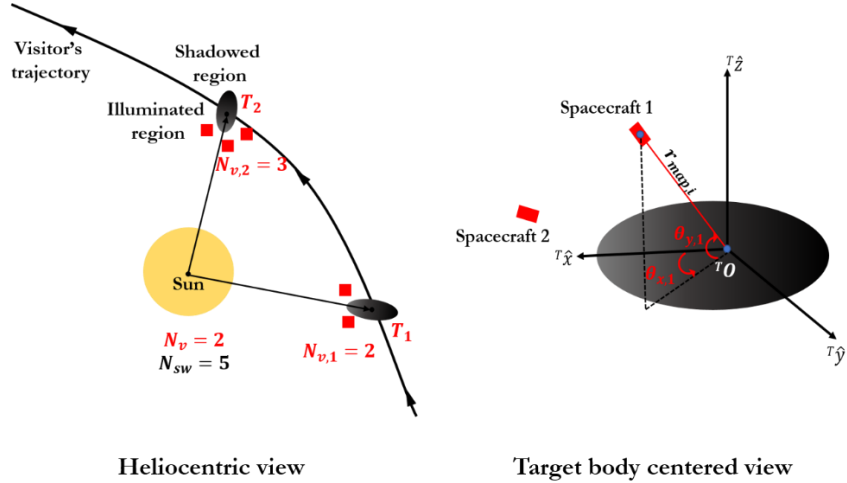

**Figure 7. A visual representation of the different design variables of the mapping swarm design problem.**

| Parameter | # spacecraft visits | # Spacecraft each visit | | Visiting time past arrival | | Spacecraft RA at visit | | Spacecraft Dec at visit | |
|---|---|---|---|---|---|---|---|---|---|
| Variable | $N_v$ | $N_{v,1}$ | ... $N_{v,N_v}$ | $T_1$ | ... $T_{N_v}$ | $\theta_{x,1}$ | ... $\theta_{x,N_{Sw}}$ | $\theta_{y,1}$ | ... $\theta_{y,N_{Sw}}$ |
| Range | Integer $[1, N_{1,max}]$ | Integer $[1, N_{2,max}]$ | | Real $[0, N_{RT}T_P]$ | | Real $[0, 2\pi]$ | | Real $[-\frac{\pi}{2}, \frac{\pi}{2}]$ | |

**Figure 8. Gene map listing the design parameters of the mapping swarm design problem.**

Once a swarm design is specified, this visual coverage obtained by the swarm is studied in order to select the optimal design. However, due to their irregular shapes, the design of a swarm can be sensitive to the orientation of the visitor. In order to address this issue, we formulate a dual sphere method as described below.

*Dual Sphere Coverage Model.* Due to their irregular shapes, the maximum $r_{T,max}$ and minimum radius $r_{T,min}$ of the target body is noted. The designed flyby should have a field of view that accounts for the maximum radius of the target body, and the spacecraft flyby location should have range enough to reach the minimum radius of the target body. Therefore, the sphere with the radius $r_{T,max}$ is referred to as the view sphere, and the sphere with radius $r_{T,min}$ is referred to as the range sphere. Both spheres are located at the center of mass of the visitor and are assumed to tumble along with the visitor. The geometry of the parameters in the dual sphere method is presented in Figure 9.

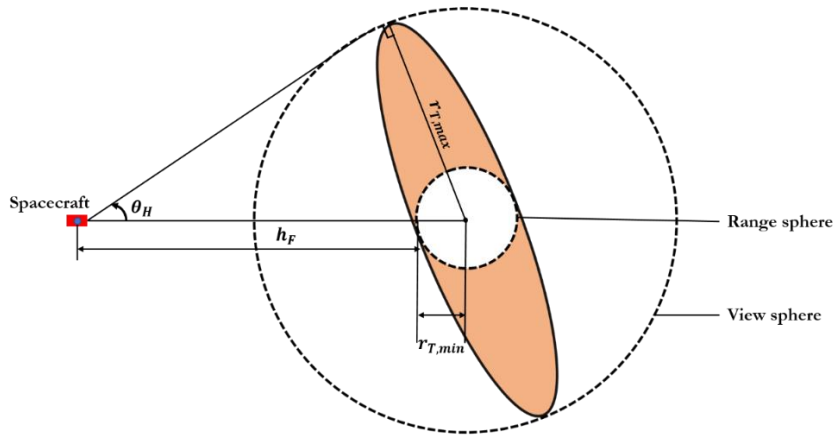

**Figure 9. The geometrical parameters used to design the spacecraft flybys using the dual sphere method.**

Using the dual sphere method, the flyby radius which accounts for the minimum flyby altitude can be computed as

$$r_{map} = h_F + r_{T,min} \qquad (13)$$

and the horizon angle that compensates for the maximum radius is computed as

$$\sin \theta_H = \left(\frac{r_{T,max}}{r_{map}}\right) \qquad (14)$$

It can be noted that Equation 14 is only valid for the cases where the spacecraft is outside the view sphere, which is the case of the designed flybys in the current work. The spacecraft FoV that compensates for this horizon angle is found from Equation 11. The sensitivity of the designs to the initial attitude is avoided by computing the coverage of the spacecraft swarm on the range and view sphere instead of the nominal shape model. We can now use the culling and clipping operations of the camera transformation[21, 31] to model the surface coverage of the model using the spacecraft swarm. It should be noted that the two spheres do not bound the portion of the nominal surface model that is mapped by the swarm. However, the complete coverage of the two spheres indicates that the nominal model is completely covered as well.

*Tumbling Dynamics.* In order to study the coverage of the swarm, the tumbling motion of the visitor must be modeled. The attitude of the visitor when tumbling is modeled using the torque-free kinematic model due to the time scale of the rendezvous operations. The modified Rodriguez parameters[32] $\bar{\sigma}$ are used to specify the visitor's attitude. The attitude kinematics of the visitor are then governed by[33]

$$\dot{\bar{\sigma}} = \frac{1}{4}[B(\bar{\sigma})]\bar{\omega}_T \qquad (15)$$

where

$$[B(\bar{\sigma})] = (1 - \sigma^2)[I] + 2[\tilde{\sigma}] + 2\bar{\sigma}\bar{\sigma}^T \qquad (16)$$

Where $[I]$ is the $3 \times 3$ identity matrix, and $\sigma$ is the magnitude of $\bar{\sigma}$, and $\bar{\omega}_T$ is the angular velocity vector. Additionally, the MRP is switched to its shadow set[33] when its magnitude exceeds unity, in order to prevent any propagation singularities, i.e.,

$$\bar{\sigma} = -\frac{\bar{\sigma}}{\sigma^2} \quad \text{if } \sigma > 1 \tag{17}$$

Additionally, the heliocentric trajectory of the visitor is propagated using a two-body model[30] given by

$$\dot{\bar{R}}_T = \bar{V}_T$$

$$\dot{\bar{V}}_T = -\frac{\mu_S}{|\bar{R}_T|^3}\bar{R}_T \tag{18}$$

Where $\mu_S$ is the gravitational parameter of the Sun. The attitude and location of the visitor during the $j^{\text{th}}$ visit is obtained by propagating Equations 15 and 18 between $T_{j-1}$ and $T_j$. The design can then be generated once the initial state of the visitor at $T_0 = 0$ is specified. Since the objects being studied are essentially spherical, the initial MRP is not important. We, therefore, specify this as a random $3 \times 1$ vector which follows the shadow switching property specified in Equation 17. Specifically, in the case of 'Oumuamua, an estimate of the spin period magnitude was known however the spin axis was not properly known. We, therefore, model the angular velocity vector as along a random spin axis, i.e.,

$$\bar{\omega}_T = \frac{2\pi}{T_P} \frac{\bar{\omega}_R}{|\bar{\omega}_R|} \tag{19}$$

Where $\bar{\omega}_R$ is a randomly generated $3 \times 1$ vector with components that are uniformly generated based on the semi-axis length along that direction, i.e. if the if semi-axis length along the $x$-axis has a length $\alpha$, the $x$-component $\bar{\omega}_R$ would be randomly distributed in $[-\alpha, \alpha]$. This characteristic is placed in order to make long axis mode spins more probably over other modes of rotation[14]. Additionally, this propagation is also used to mark the faces of the shape model that will be illuminated by the Sun and can consequently be mapped by a Swarm. The shape model is subjected to a culling operation[21, 23] using the heliocentric position vector of the visitor to identify the illuminated faces at a given instant.

*Spacecraft Attitude Behaviors.* The spacecraft in the swarm are assumed to follow a Class 2 swarm architecture[21], where the spacecraft coordinate sensing and communications. The swarm is assumed to have an Observing spacecraft and a Leader spacecraft. All spacecraft are assumed to track the line of sight (LoS) with respect to a target[22]. If the distance from the spacecraft to the target body $r_{Ts}$ falls inside the imaging radius, i.e., $r_{map} + \Delta h$, the spacecraft will point their cameras toward the LoS with respect to the target. If the Observing spacecraft are outside the imaging region, their communication axis will track the LoS with respect to the Leader spacecraft. When the Leader spacecraft is outside the imaging region its communication axis will track the LoS with respect to Earth. The leader spacecraft in the swarm is selected based on the same criteria described in Reference 23. The triad algorithm is used to define the reference attitude for the swarm using the LoS axis, and the relative velocity vector with respect to the target. In order to determine the heliocentric velocity of the spacecraft during the rendezvous, the boundary value algorithm described in Reference 24 is used. This allows us to compute the relative velocity vector in order to compute the reference attitude. Additionally, the swarm is assumed to have no collisions either among itself or with the target. The collisions with the target body are checked by verifying that $r_{Ts} > r_{T,max}$ and the inter-spacecraft collisions are checked by ensuring that distance between spacecraft $i$ and spacecraft $j$ does not fall below a collision radius $r_{col}$ during the entire propagation period, and for all $i, j \leq N_{Sw}$.

*Figure of Merit.* Due to the uncertainties arising from the attitude and tumbling of the simulated visitor, each design generated to $N_{mon,2}$ Monte-Carlo simulations with randomly generated initial attitude and spin-axis vectors. For each simulated design, the swarm coverage after all passes is noted for both the view and the range spheres. The design is said to be successful if its minimum average surface area mapped on both the spheres $\min(\langle P_{View}\rangle_{N_{mon,2}}, \langle P_{Range}\rangle_{N_{mon,2}})$ exceed the requirement $P_{Cov}$. Where $\langle P_{View}\rangle_{N_{mon,2}}$ and $\langle P_{Range}\rangle_{N_{mon,2}}$ denote the percentage of the surface area mapped on the view and range spheres respectively when subjected to $N_{mon,2}$ simulations.

*Automated Mapping Swarm Design.* The automated swarm design problem that is solved by the Swarm Designer module of IDEAS can now be specified as

$$\min J_{sw} = N_{Sw} = \sum_{j=1}^{N_v} N_{v,j}$$

such that

$$\min(\langle P_{View}\rangle_{N_{mon,2}}, \langle P_{Range}\rangle_{N_{mon,2}}) > P_{Cov}$$

$$\text{Collsions} = \text{False} \tag{20}$$

Additionally, a linear constraint that ensures that visiting times of the swarms are designed in ascending order is placed. This describes the methodologies used in the current work to design the constellation, trajectories, and visual mapping swarm to explore the interstellar visitors. The next section will demonstrate the algorithms described in the current work using numerical simulations of a detection constellation and a rendezvous swarm.

**Table 1. Input parameters to the detection swarm design problem.**

| Parameter | Value |
| --- | --- |
| Radius range | [0.7, 1.5] AU |
| $P_r$ | 85 % |
| $P_{Op}$ | 90 % |
| Half-FoV | 30 deg |
| $N_{mon1}$ | 10,000 |
| $N_{orb}$ range | [1,20] |
| $N_{rot}$ range | [1,20] |
| $N_p$ range | [1,25] |
| $N_{sp}$ range | [1,25] |
| $F$ range | [1,4] |

**NUMERICAL SIMULATIONS**

In this section, we perform simulation case studies to demonstrate the algorithms discussed in the previous section. The design optimization problems formulated in Equations 6, 8, and 20 are solved using a mixed-integer constrained genetic algorithm optimization algorithm[34] embedded in MATLAB.

We begin by designing the detection constellation. Then assuming there was enough forewarning time, i.e., if detected inside the boundary limits of the constellation design a visual mapping swarm to explore 1I/'Oumuamua.

**Detection Swarm Design**

As mentioned above, the detection swarm will be designed to detect events within the radial region $[0.7, 1.5]$ AU. The constellation is required to meet a minimum detection efficiency of 85 %, with an operational percentage of 90 %. Each design is subjected to 10,000 randomly generated visitor events. The input parameters to the optimization problem in Equation 6 are presented in Table 1.

*Optimization.* Each genetic optimization trial of the detection swarm design problem in Equation 6 explored a population of 100 designs per generation. The optimizer was able to converge to a solution within about 30 generations. However, the local optimality was observed when the same solution was stalled as the optimal solution for the subsequent 20 generations. The solution was verified by running the optimizer 5 times, where all trials observed to converge to the same solution. The optimizer convergence for the 5 trials showing the evolution of the run-averaged means and best solution at each generation along with the selected optimal detection constellation gene are presented in Figure 10.

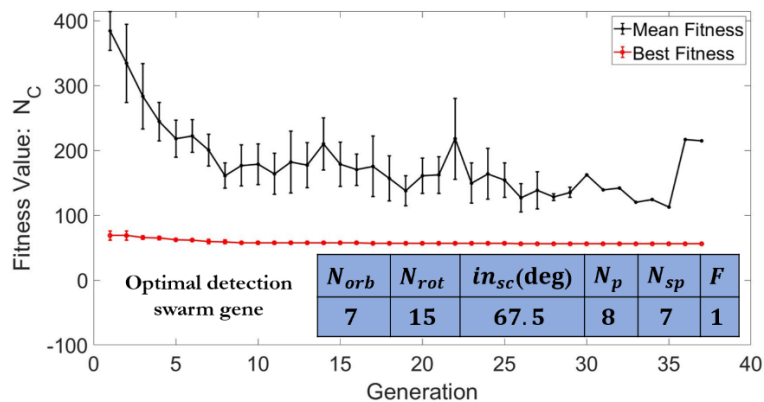

**Figure 10. The evolution of the mean and best designs across different generations of the optimal detection swarm design problem, along with the selected optimal gene.**

*Performance.* As noted from Figure 10, the optimal detection swarm is a $67.5: 56/8/1$ Walker-Delta constellation. The constellation has 7 spacecraft each plane, with 8 such orbital planes, where each neighboring plane is phased $\frac{\pi}{56}$ rad counterclockwise with respect to its immediate western adjacent orbital plane. The constellation has orbits that repeat their ground track in a $7:15$ RGT pattern. The constellation design was noted to have a detection efficiency of 85.6% for a random sample of 10,000 simulated visitor events, indicating that nearly 856 of these simulated events were detected by at least 2 spacecraft in the swarm, while 90 % of its population was operational, and the specified pointing constraints were met. An example event detection of the constellation is visualized in Figure 11.

**Trajectory Design**

As a notional case, we design a flyby mission to the visitor 1I/'Oumuamua assuming it was detected in the $[0.7, 1.5]$ AU range. As mentioned above the trajectories are designed as Lambert-arc search problem from the Earth to the visitor.

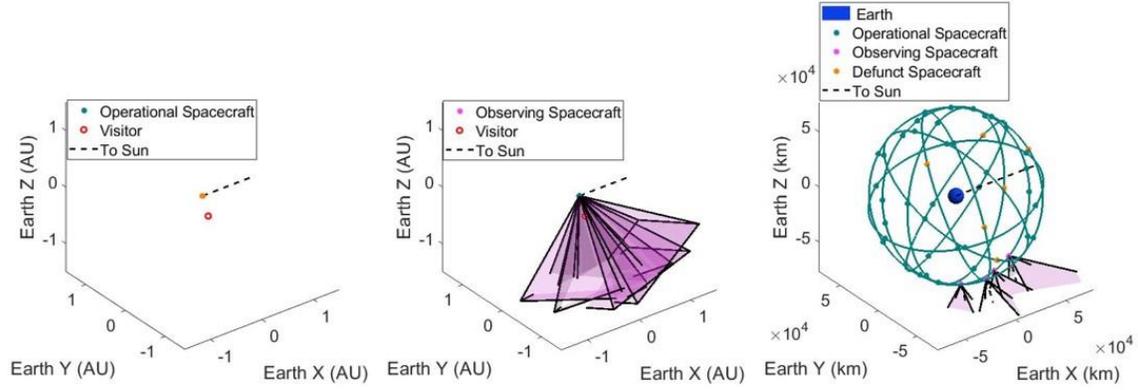

**Figure 11. The simulated performance of the designed detection swarm, where a randomly generated visitor event is observed by at least 2 spacecraft in the constellation.**

It is noted from the ephemeris of 'Oumuamua[5] noted on 23 Nov 2017, the visitor was already on its outbound heliocentric trajectory at an estimated distance of about 1.44 AU from the Earth. Using the orbital element phasing algorithm[30] the ephemeris of the visitor on 1 Aug 2017 are noted, where the visitor was still on its incoming asymptote and was at a distance of 1.34 AU from the Earth. The Lambert arcs were then constructed from this lower bound to the specified arrival dates. All Lambert problems were solved as a 0-rev min-$C_3$ arcs using a fast Lambert solver algorithm[35]. The parameters input to the trajectory design problem are presented in Table 2.

**Table 2. Input parameters to the trajectory design problem.**

| Parameter | Value |
| --- | --- |
| $D_L$ range | 1 Aug 2017 − 31 Dec 2021 |
| $ToF$ range | 50 − 1000 days |
| $C_{3,max}$ | 200 km²/s² |
| $v_{\infty,max}$ | 100 km/s |

The trajectory design problem in Equation 8 was solved using the genetic algorithm optimizer. The algorithm was able to converge to an optimal solution with in the first 2 − 3 generations of the genetic algorithm where each gene spanned a population of 2650 designs. The selected result of the trajectory problem is presented in Figure 12 along with the $C_{3,E}$ and $v_{\infty,T}$ porkchop contours. The contours highlighted in Figure 12 correspond to the ranges marked by the bounds specified in Table 2. As noted from Figure 12, the entire set of feasible trajectories for launch are located to the left of the date for which the ephemeris is present, which again confirms that optimal missions can be designed if there is enough forewarning time. The trajectory should be deployed from the Earth on 1 Aug 2017 and would arrive at 'Oumuamua on 17 Oct 2017, with a total cruise time of about 77.7 days. The $C_3$ launch energy is noted to be 18.3 km²/s² which is indeed capable by most launch providers[36]. The excess velocity magnitude at rendezvous with 'Oumuamua will be 55.5 km/s. It is acknowledging here that this can indeed be challenging for the spacecraft to rendezvous with such high speeds. Other techniques such as gravity assist, or low thrust propulsion methods[18] can be used to mitigate this magnitude while trading this with the time of flight. The selected heliocentric trajectory of the swarm is presented in Figure 13. It can be noted from Figure 13 that the rendezvous will occur in a plane closer to the Earth's orbit, after the perihelion passage of 'Oumuamua.

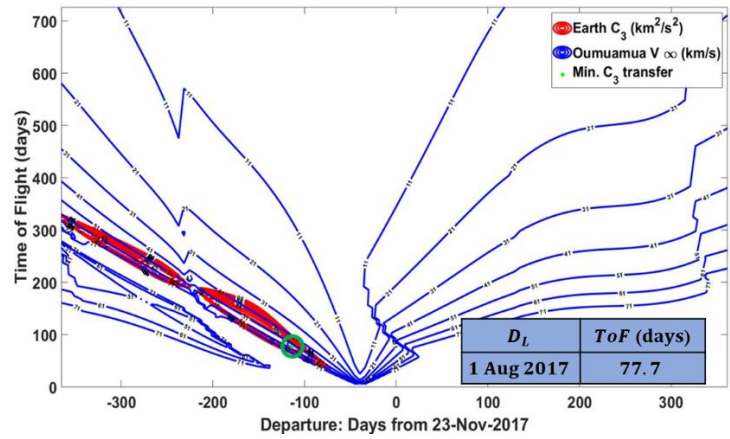

**Figure 12. The Earth-Visitor porkchop plot showing the selected optimal trajectory and its design gene.**

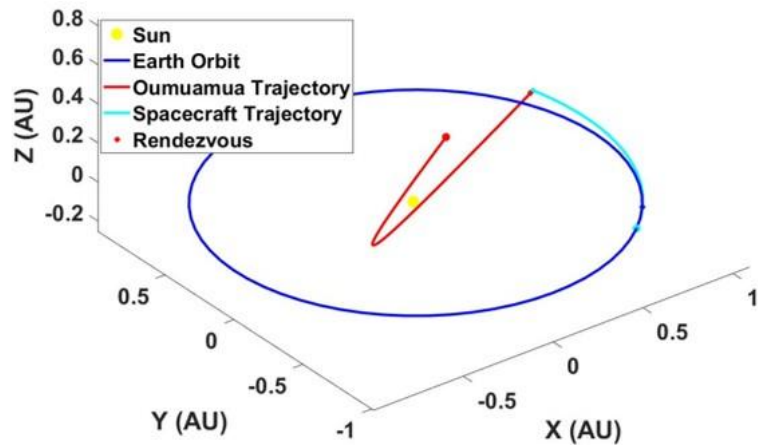

**Figure 13. The designed optimal heliocentric trajectory of the swarm originating from the Earth and traveling to the 'Oumuamua.**

**Visual Mapping Swarm Design**

Upon rendezvous, the objective is to generate global surface maps of 'Oumuamua at a minimum ground resolution of 10 cm and a minimum elevation angle of 5 deg. The swarm will be expected to map at least 85 % of 'Oumuamua's surface. The visual mapping swarm is then designed by solving the optimization problem presented in Equation 20, by using the dual sphere coverage method. A triaxial ellipsoid model with semi-axis $230 \times 35 \times 35$ m is used to model the nominal shape of 'Oumuamua. Additionally, a uniformly random perturbation with amplitude $\pm 1\%$ is added to the nominal ellipsoid vertices to generate surface irregularities on the shape model. The view sphere is modeled as a sphere of radius 230 m, while the range sphere is modeled as a sphere of radius 35 m. The shape models are then generated as triangular comprised of about 5100 triangles each. The shape models used in the current work are presented in Figure 14. The models are subjected to a constant tumbling spin rate magnitude with a period of 8.14 hrs, however, the spin axis is randomized during each simulation trial described by Equation 19.

*Optimization.* Each genetic optimization trial of the mapping swarm design problem explored a population of 50 designs per generation. The simulation runs were set up to run a maximum of 50 generations, with a stalling stop criterion 10 generations. The parameters used to solve the mapping swarm design problem are presented in Table 3. The swarm optimization problem is solved using a genetic algorithm optimization scheme. The result of the optimization run showing the evolution of the mean and best swarm design across the design generations is presented in Figure 15. As seen in Figure 15, the genetic algorithm was able to locate an optimal solution within 30 generations, after which the best design stalled for the set criterion of 10 generations.

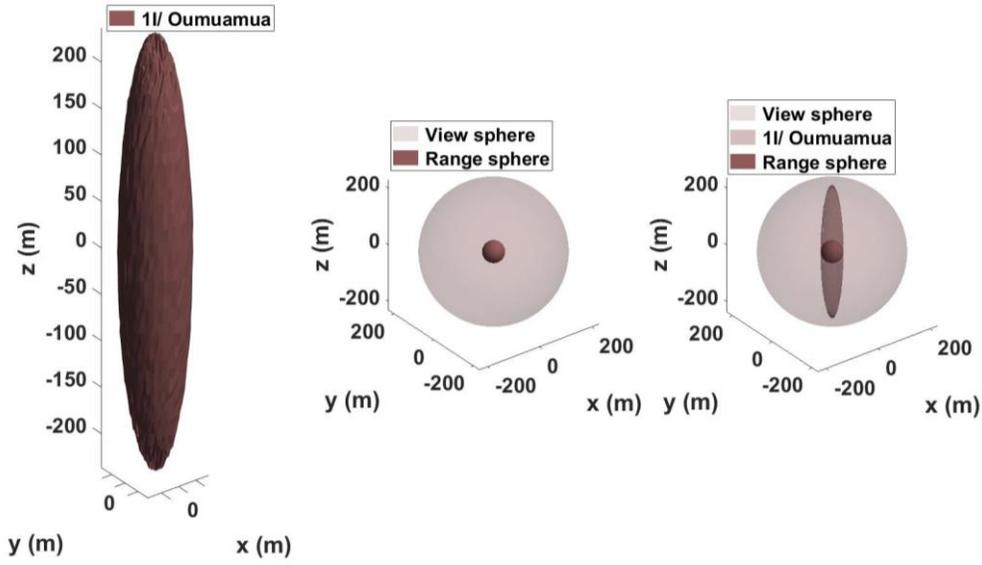

**Figure 14. The nominal and dual sphere of 1I 'Oumuamua used in the current work to study the swarm coverage.**

*Performance.* As noted in Figure 15, the optimal swarm consists of a total of 7 spacecraft that will visit 'Oumuamua at 4 locations. The first simulated mapping operation visit of the swarm is presented in Figure 16.

**Table 3. Parameters corresponding to the mapping swarm design problem.**

| Parameter | Value |
|---|---|
| $x_{Des}$ | 10 cm |
| $\varepsilon_{Des}$ | 5 deg |
| $P_{Cov}$ | 85 % |
| $D_{sc}$ | 8 cm |
| $h_{max}$ | 11.5 km |
| $\Delta h$ | 5 km |
| $N_{mon,2}$ | 250 |
| $N_v$ range | [1,5] |
| $N_{vj}$ range | [1,5] |
| $N_{RT}$ | 3 |
| $T_P$ | 8.14 hrs |
| $r_{col}$ | 1 m |

The designed swarm is expected to cover 85.9% of the view sphere and 86.0 % of the range sphere. The standard deviation of the coverage for both the spheres was noted to be around 12.5% over a random 250 attitude configurations of the shape model. It is noted here that certain spin states of the visitor can limit the coverage. For example, when the spin axis is closely aligned with or against the Sun vector, the eclipsed portion is virtually inaccessible during the rendezvous. For this reason, the mean coverage of the spheres is taken as the figure of merit. The final accumulated coverage of all the spheres during a simulated trial is presented in Figure 17. During this trial, the swarm was able to map nearly 98.8 % of the view sphere, 99.3 % of the range sphere, and 97.0 % of the nominal shape model, as indicated in Figure 17.

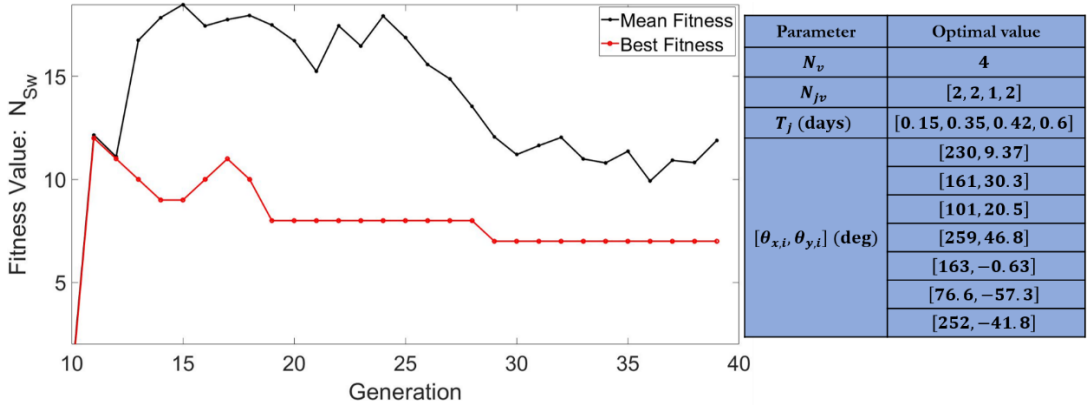

**Figure 15. The evolution of the mean and best designs across different generations of the optimal mapping swarm design problem, along with the selected optimal gene.**

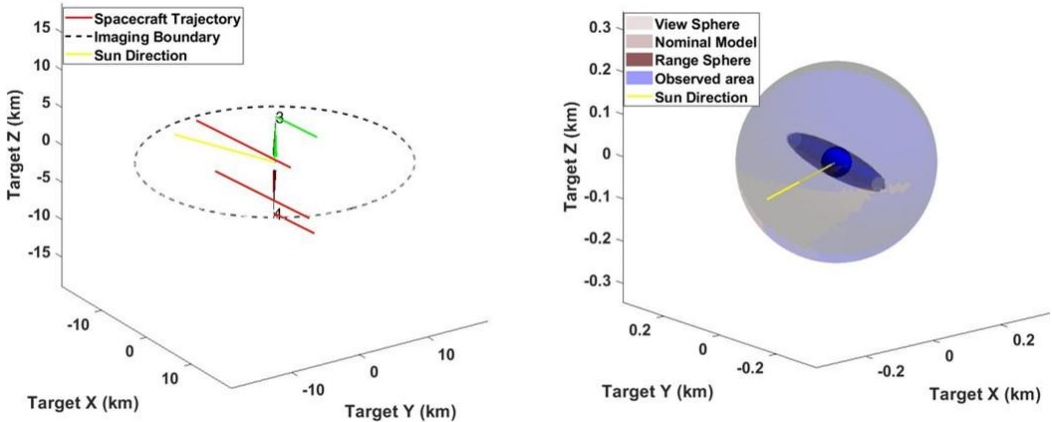

**Figure 16. Visualization of a simulated mapping visit generated by the optimizer, showing the spacecraft trajectories (left), and the instantaneous coverage on all the shape models (right).**

**Discussion**

The current work highlights certain key points that are factored into the designing swarm missions to interstellar visitors. First, the arrival of the interstellar visitors is a random process, therefore deterministic tools cannot be used to design detection constellations. Secondly, as demonstrated in the current work, the detection swarm design, mapping swarm, and their trajectories are independent: The mapping swarm design is based on the detection range of the system, which provides the required forewarning time. Decoupled designs of the swarms can produce optimal results but might not be compatible with one another. The IDEAS architecture is specially designed to handle such inter-related problems. Finally, it is acknowledged here that while the simulated missions were indeed able to meet the design criterion, the performance of a real time swarm mission does not necessarily have to yield such optimal results.

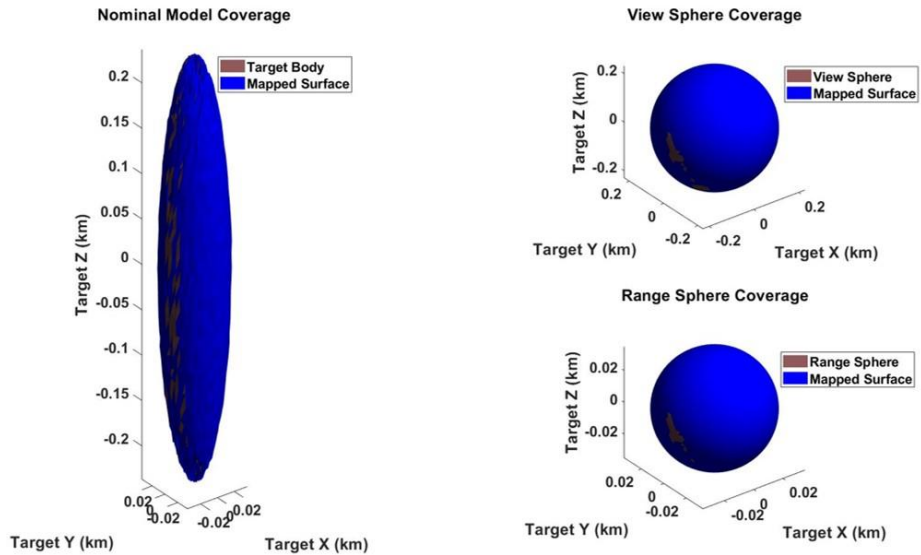

**Figure 17. Cumulative surface coverage of the nominal shape model (left), and its corresponding dual spheres (right) noted after the flybys of all spacecraft in the swarm.**

These limitations can arise from spacecraft limitations such as pointing accuracy, or modeling uncertainties such as unaccounted dynamics. For this reason, statistical tools are developed to place soft bounds on the results. Additionally, developing the Spacecraft Designer module of IDEAS to maintain a database of state-of-the-art spacecraft subsystems can also improve the fidelity of these results.

The following are the important contributions of the current work: This work developed statistical algorithms to design an interstellar visitor detection constellation to provide enough forewarning time. Following this, we developed the dual sphere method to mitigate the sensitivity of spacecraft coverage to the initial attitude of the visitor. Finally, we developed statistical algorithms to model spacecraft coverage near a tumbling small body, whose spin axis is not properly known.

**CONCLUSION**

This work developed statistical algorithms to design swarm missions to explore interstellar visitors using the IDEAS architecture. We began with the design of a detection swarm that meets a required detection criterion. Then, assuming that such a swarm allows enough forewarning time, a notional swarm flyby mapping mission to visually map these visitors was presented. The design of the heliocentric trajectories of the swarm was presented, followed by designing the rendezvous swarm that meets its surface coverage requirement. The visitor in this work is assumed to undergoing a tumbling motion along a random spin axis. In order to reduce the sensitivity of the designs to the initial attitude, a dual sphere method was presented to study the visual coverage of the swarms. The coverage was then noted from a Monte-Carlo simulation of random initial attitude configurations of the visitor. Finally, the algorithms described in the current work were demonstrated through numerical simulations of designing a detection swarm, and a rendezvous mission to 'Oumuamua assuming enough forewarning time.

While the current work focused on designing rendezvous missions assuming heliocentric dynamics of both spacecraft and the visitor, future work using IDEAS will focus on implementing high fidelity dynamical models. Specifically, perturbations such as solar radiation pressure, and third body gravitation will be modeled into the design. Such perturbations can be used to model probabilistic maneuvers to correct for errors from nominal dynamics. Additionally, while the current work focused on developing theoretical missions, future work on IDEAS will be used to develop hardware testbeds that demonstrate the optimal swarm operations in multi-agent autonomous systems. These studies can be used to validate the designs and demonstrate the flow of information when designing swarm missions, thus enabling IDEAS to be an end-to-end tool to design swarm missions to explore the small bodies in the solar system.

# REFERENCES

[1] Schneider, J., 2017. Is 1I/2017 U1 really of interstellar origin? *Research Notes of the AAS*, *1*(1), p.18.

[2] Bannister, M.T., Bhandare, A., Dybczynski, P.A., Fitzsimmons, A., Guilbert-Lepoutre, A., Jedicke, R., Knight, M.M., Meech, K.J., McNeill, A., Pfalzner, S. and Raymond, S.N., 2019. The natural history of 'Oumuamua. *Nature Astronomy*, *3*, pp.594-602.

[3] Ye, Quan-Zhi, Qicheng Zhang, Michael SP Kelley, and Peter G. Brown. "1I/2017 U1 (Oumuamua) is hot: imaging, spectroscopy, and search of meteor activity." *arXiv preprint arXiv:1711.02320* (2017).

[4] Guzik, P., Drahus, M., Rusek, K., Waniak, W., Cannizzaro, G. and Pastor-Marazuela, I., 2019. Interstellar comet 2I/Borisov. *arXiv preprint arXiv:1909.05851*.

[5] Johnston, R. (2019). *1I 'Oumuamua (2017 U1): The first interstellar object within our solar system*. [online]: http://www.johnstonsarchive.net/astro/oumuamua.html [Accessed 21 Jan. 2020].

[6] De la Fuente Marcos, C., and de la Fuente Marcos, R., 2017. Pole, Pericenter, and Nodes of the Interstellar Minor Body A/2017 U1. *Research Notes of the American Astronomical Society*, *1*(1).

[7] Do, A., Tucker, M.A. and Tonry, J., 2018. Interstellar interlopers: number density and origin of 'Oumuamua-like objects. *The Astrophysical Journal Letters*, *855*(1), p.L10.

[8] Laughlin, G. and Batygin, K., 2017. On the Consequences of the Detection of an Interstellar Asteroid. *Research Notes of the AAS*, *1*(1), p.43.

[9] Ferrin, I. and Zuluaga, J., 2017. 1I/2017 U1 (Oumuamua) Might Be A Cometary Nucleus. *arXiv preprint arXiv:1711.07535*.

[10] Jewitt, D. and Luu, J., 2019. Initial Characterization of Interstellar Comet 2I/2019 Q4 (Borisov). *The Astrophysical Journal Letters*, *886*(2), p.L29.

[11] Jewitt, D., Luu, J., Rajagopal, J., Kotulla, R., Ridgway, S., Liu, W. and Augusteijn, T., 2017. Interstellar Interloper 1I/2017 U1: Observations from the NOT and WIYN Telescopes. *The Astrophysical Journal Letters*, *850*(L36), p.7pp.

[12] Fraser, W.C., Pravec, P., Fitzsimmons, A., Lacerda, P., Bannister, M.T., Snodgrass, C. and Smolić, I., 2018. The tumbling rotational state of 1I/'Oumuamua. *Nature Astronomy*, *2*(5), pp.383-386.

[13] Bolin, B.T., Weaver, H.A., Fernandez, Y.R., Lisse, C.M., Huppenkothen, D., Jones, R.L., Jurić, M., Moeyens, J., Schambeau, C.A., Slater, C.T. and Ivezić, Ž., 2017. APO time-resolved color photometry of highly elongated interstellar object 1I/'Oumuamua. *The Astrophysical Journal Letters*, *852*(1), p.L2.

[14] Scheeres, D.J., 2016. *Orbital motion in strongly perturbed environments: applications to asteroid, comet, and planetary satellite orbiters*. Springer.

[15] Drahus, M., Guzik, P., Waniak, W., Handzlik, B., Kurowski, S. and Xu, S., 2018. Tumbling motion of 1I/ 'Oumuamua and its implications for the body's distant past. *Nature Astronomy*, *2*(5), pp.407-412.

[16] Castillo-Rogez, J.C., Landau, D., Chung, S.J., and Meech, K., 2019. Approach to exploring interstellar objects and long-period comets.

[17] Hibberd, A., Hein, A.M., and Eubanks, T.M., 2020. Project Lyra: Catching 1I/'Oumuamua–Mission Opportunities After 2024. *Acta Astronautica*.

[18] Hein, A.M., Perakis, N., Eubanks, T.M., Hibberd, A., Crowl, A., Hayward, K., Kennedy III, R.G. and Osborne, R., 2019. Project Lyra: Sending a spacecraft to 1I/'Oumuamua (former A/2017 U1), the interstellar asteroid. *Acta Astronautica*, *161*, pp.552-561.

[19] Brophy, J., Polk, J., Alkalai, L., Nesmith, B., Grandidier, J. and Lubin, P., 2018. A Breakthrough Propulsion Architecture for Interstellar Precursor Missions: Phase I Final Report.

[20] Hibberd, A., Perakis, N., and Hein, A.M., 2019. Sending a Spacecraft to Interstellar Comet C/2019 Q4 (Borisov). *arXiv preprint arXiv:1909.06348*.

[21] Nallapu, R., and Thangavelautham, J., 2019, March. Attitude Control of Spacecraft Swarms for Visual Mapping of Planetary Bodies. In *2019 IEEE Aerospace Conference*. IEEE.

[22] Nallapu, R., and Thangavelautham, J., "Spacecraft Swarm Attitude Control for Small Body Surface Observation," *In 2019 Advances in the Astronautical Sciences (AAS) GNC Conference Proceedings*, AAS-GNC, 2019.

[23] Nallapu, R., and Thangavelautham, J., "*Towards End-To-End Design of Spacecraft Swarms for Small-Body Reconnaissance*," Proceedings of the 70th International Astronautical Congress, IAF, Washington D.C.

[24] Nallapu, R., and Thangavelautham, J., 2020. Design of Spacecraft Swarm Flybys for Planetary Moon Exploration. In *AIAA Scitech 2020 Forum* (p. 0954).

[25] Wertz, J.R., 2001. Mission geometry: orbit and constellation design and management: spacecraft orbit and attitude systems. *Mission geometry: orbit and constellation design and management: spacecraft orbit and attitude systems/James R. Wertz. El Segundo, CA; Boston: Microcosm: Kluwer Academic Publishers, 2001. Space technology library; 13*.

[26] Wertz, J.R., Everett, D.F. and Puschell, J.J., 2011. *Space mission engineering: the new SMAD*. Microcosm Press.

[27] Nallapu, R., Kalita, H. and Thangavelautham, J., 2018. On-Orbit Meteor Impact Monitoring Using CubeSat Swarms.

[28] Nallapu, R., and Thangavelautham, J., "Cooperative Multi-spacecraft Observation of Incoming Space Threats", 2019 Advanced Maui Optical Science (AMOS) Proceedings, AMOS Tech.,2019.

[29] Nallapu, R., Vance L.D., Xu Y., and Thangavelautham, J., 2020 Automated Design Architecture for Lunar Constellations. *2020 IEEE Aerospace Conference, 2020*.

[30] Vallado, D.A. & McClain, W.D., 2013. *Fundamentals of astrodynamics and applications* 4th ed., Hawthorne, CA: Microcosm Press.

[31] Sobel, I., 1972. *Camera models and machine perception* (No. CS Technion report CS0016). Computer Science Department, Technion.

[32] Terzakis, G., Lourakis, M. and Ait-Boudaoud, D., 2018. Modified Rodrigues Parameters: an efficient representation of orientation in 3D vision and graphics. *Journal of Mathematical Imaging and Vision*, *60*(3), pp.422-442.

[33] Schaub, et al. *Analytical Mechanics of Space Systems*. 3rd ed., American Institute of Aeronautics and Astronautics, 2014.

[34] Conn, A.R., Gould, N.I. and Toint, P., 1991. A globally convergent augmented Lagrangian algorithm for optimization with general constraints and simple bounds. *SIAM Journal on Numerical Analysis*, *28*(2), pp.545-572.

[35] Russell, R.P., 2019. On the solution to every lambert problem. *Celestial Mechanics and Dynamical Astronomy*, *131*(11), p.50.

[36] Guo, Y. and Farquhar, R.W., 2008. New Horizons mission design. *Space science reviews*, *140*(1-4), pp.49-74.